\documentstyle[leqno,twoside,11pt,ltexproc]{article}
\begin{document}
\cleardoublepage
\pagestyle{myheadings}

\title{Observing Binary Inspiral with LIGO\thanks{To appear in Proceedings
of the Lanczos International Centenary. Research
supported by grants from NASA (NAGW-2936), the NSF (PHY 9308728),
and the Alfred P.\ Sloan Foundation.}}
\author{Lee Samuel Finn\thanks{Alfred P.\ Sloan Research Fellow,
Department of Physics and Astronomy, Northwestern University.}}
\date{}
\maketitle
\markboth{L.\ S.\ Finn}{Observing Binary Inspiral with LIGO}
\pagenumbering{arabic}
\begin{abstract}
Gravitational radiation from a binary neutron star or black hole
system leads to orbital decay and the eventual coalescence of the
binary's components. During the last several minutes before the
binary components coalesce, the radiation will enter the
bandwidth of the United States Laser Inteferometer
Gravitational-wave Observatory (LIGO) and the French/Italian
VIRGO gravitational radiation detector.  The combination of
detector sensitivity, signal strength, and source density and
distribution all point to binary inspiral as the most likely
candidate for observation among all the anticipated sources of
gravitational radiation for LIGO/VIRGO. Here I review briefly
some of the questions that are posed to theorists by the
impending observation of binary inspiral.
\end{abstract}

\section{Introduction}

The questions raised by the impending observation of
gravitational radiation can be divided into three broad classes:
data analysis, ``anticipatory'' astrophysics, and
``contributory'' astrophysics. By data analysis I mean the
techniques that convert the output of a detector system into
statements regarding the likelihood that a source of radiation
has been observed and characterizing the nature of the source. In
order to carry out the data analysis we must draw upon our
knowledge of the (astro)physics of the sources. The questions so
posed ({\em e.g.,\/} what are the waveforms, where are the
sources located, how many sources are there?)  are those I
categorize as anticipatory astrophysics.  Finally, the direct
detection of gravitational radiation will allow us to add to our
understanding of the physics of gravitation ({\em e.g.,\/} tests
of general relativity), sources, and astrophysics generally. I
categorize the astrophysics questions that can be posed of the
observations as contributory astrophysics.

Questions of anticipatory and contributory astrophysics are
defined by what is needed for data analysis (in the first case)
and what it is possible to learn given the detector output (in
the second). Consequently, our first goal must be to understand
the data analysis process well enough to define what we need to
carry it out and what it is possible to learn given our detection
system.

The output of a detector is a time series $g(t)$. In any given
time interval, $g(t)$ consists either of detector noise alone or
detector noise in combination with a gravitational wave signal:
\begin{equation}
g(t) = \left\{\begin{array}{ll}
n(t)&\mbox{detector noise;}\\
n(t)+s(t;\mbox{\boldmath$\mu$})&\mbox{noise + signal.}
\end{array}\right.
\end{equation}
I assume here that the detector response to a source of
radiation superposes linearly with the detector noise.  The
detector response to the gravitational radiation from a
particular class of sources, such as inspiraling binaries, is
represented by $s(t;\mbox{\boldmath$\mu$})$. Finally, the
(vector-valued) parameter {\boldmath$\mu$} fully characterizes
the radiation from the particular source ({\em e.g.,\/} the
distance to the source, the masses of the binary components, {\em
etc.}). {\em The goal of data analysis is to determine the
conditional probability density $P[s(\mbox{\boldmath$\mu$})|g]$ that
the signal $s(\mbox{\boldmath$\mu$})$ is present given the detector
output $g$.}

Using Bayes law of conditional probabilities we can express
$P[s(\mbox{\boldmath$\mu$})|g]$ in terms of $g(t)$, the signal model
$s(\mbox{\boldmath$\mu$})$, the power spectral density of the detector
noise $S_n(f)$, and two {\em a priori\/} probabilities: $P(s)$,
the {\em a priori\/} probability that a signal of the class
represented by $s$ is present in $g(t)$, and
$P(\mbox{\boldmath$\mu$}|s)$, the {\em a priori\/} probability that a
random source of type $s$ is characterized by {\boldmath$\mu$}
(for details, see \cite{finn92a} and references therein).

\section{Anticipatory astrophysics for binary inspiral}

The anticipatory astrophysics questions that must be addressed
are those that lead to the determination of the {\em a priori\/}
probabilities $P(s)$ and $P(\mbox{\boldmath$\mu$}|s)$. In the case of
binary inspiral, $s(\mbox{\boldmath$\mu$})$ represents the detector
response to the binary inspiral waveform; thus, the first
challenge to theorists is to determine the inspiral radiation
waveform $h(\mbox{\boldmath$\mu$})$.

The waveform $h(\mbox{\boldmath$\mu$})$ is estimated via a
post-Newtonian (weak field, slow-motion) perturbation expansion
of the solution to the field equations of general relativity. The
determination of the waveform for coalescing binary systems to
greater and greater levels of accuracy has been the focus of Will
and collaborators (cf.\ \cite{wiseman93a} and references
therein). An unsettled question is how far the perturbation
expansion must be taken before increasing refinements of the
waveform are beyond the capability of the detector to
distinguish. At present it appears that physical effects that
enter at the post${}^{5/2}$Newtonian level will be discerned with
reasonable probability using the advanced LIGO instrumentation
\cite{cutler94a}; consequently, the perturbation expansion needs
to be carried out to a higher level in order to take the greatest
advantage of binary inspiral observations.

To determine $P(s)$ we are called upon to estimate the rate that
binary inspiral occurs throughout the universe. Estimates of this
rate are largely empirical and based upon our observation of the
very few observed binary pulsar systems in our own galaxy and its
globular clusters; consequently, on top of the small number
statistics of the observed binaries, we must account also for the
selection effects that lead to the {\em observed\/} binary pulsar
systems as an unknown fraction of the total number of binary {\em
pulsar\/} systems, that are (in turn) an unknown fraction of the
total number of binary {\em neutron star\/} systems. To turn this
rate density into a rate we must integrate over the past light
cone of the detector. Since LIGO will be sensitive to inspirals
at cosmological distances, we are also forced to confront our
uncertain knowledge of source evolution and the kinematics of the
expanding universe \cite{finn93a,chernoff93a}.  Thus, our
extrapolation from the local neighborhood is very uncertain;
presently, our best estimate of the rate density of neutron star
- neutron star inspiral is
$8\times10^{-8}\mbox{Mpc}^{-3}\mbox{yr}^{-1}$
\cite{narayan91a,phinney91a}.

Finally, $P(\mbox{\boldmath$\mu$}|s)$ represents a summary of our
understanding of the range and distribution of neutron star
masses (and, to a lesser extent, spin) as they occur in binary
systems, as well as the space density of coalescing binaries as a
function of luminosity distance and redshift. Thus we are forced
to confront and summarize our rather thin knowledge of the
distribution of neutron star masses.

One question that can be addressed without knowledge of either
$P(\mbox{\boldmath$\mu$}|s)$ or $P(s)$ is the anticipated {\em
precision\/} of {\boldmath$\mu$} measurements {\em in the limit
of high signal-to-noise ratio $\rho$\/} \cite{finn92a,finn93a}.
One conclusion worthy of note is that the combination of binary
component masses and system redshift ${\cal
M}\equiv\mu^{3/5}M^{2/5}(1+z)$ can be determined to a phenomenal
precision: better than a part in $10^{-2}/\rho$ for $\rho\geq10$
\cite{finn93a,cutler94a}.

\section{Contributory astrophysics}

Observation of binary inspiral in gravitational radiation is an
especially exciting prospect. {\em Exactly because the
astrophysical issues that we are forced to address in preparing
for the data analysis touch upon areas of deep uncertainty and
ignorance, binary inspiral observations will yield an incredible
wealth of new knowledge.} The impending observations will measure
the rate of expansion of the universe entirely independent of any
existing techniques and free of the uncertainties that plague the
cosmic distance ladder \cite{schutz86a,chernoff93a,markovic93a}.
Neutron star mass determinations, now possible with high
precision only for the handful of binary pulsar systems where
relativistic effects are important, will be made for every binary
inspiral observed --- and, corresponding to the best estimate
binary inspiral rate density, $\sim100\,\mbox{yr}^{-1}$ neutron
star masses will be measured by the advanced LIGO instruments
with signal-to-noise ratio greater than 8. These two examples
scratch only the surface of the possibilities inherent in LIGO
observations.

\section{Conclusions}

The study of how the direct detection of gravitational radiation
can be exploited to yield a greater understanding of the universe
has not yet received the attention it deserves: the technical
challenges of detecting the radiation and analyzing the data
until recently have consumed most of the effort of the small
community of relativists --- both theoretical and experimental
--- who have been active in confronting that challenge. Binary
inspiral is the most likely of the anticipated sources to be
observed --- given our current understanding of the
universe. Other anticipated sources may be observed first; the
more adventurous among us expect earlier observations of entirely
unanticipated sources.  Regardless, we can all look forward with
confidence to the direct detection of gravitational radiation,
and beyond that to the greater knowledge of the universe that
those observations will bring.


\end{document}